\documentstyle[preprint,aps,psfig,amsmath]{revtex}

\begin{document}

\begin{center} 
Effects of T=0 two body matrix elements on M1 and Gamow-Teller transitions: isospin decomposition
\medskip 
\\ 
Shadow J.Q. Robinson$^{a}$ and Larry Zamick$^{a,b}$

\noindent a) Department of Physics and Astronomy,
Rutgers University, Piscataway, \\New Jersey  08855

\noindent b)  TRIUMF,4004 Wesbrook Mall, Vancouver, British 
Columbia, \\Canada, V6T 2A3

\end{center}

\bigskip
\begin{abstract}
We perform calculations for M1 transitions and allowed Gamow Teller
(GT) transitions in the even-even Titanium isotopes - $^{44}$Ti,
$^{46}$Ti, and $^{48}$Ti.  We first do calculations with the FPD6
interaction. Then to study the effect of T=0 matrix elements on
the M1 and GT rates we introduce a second interaction in which all the
T=0 matrix elements are set equal to zero and a third in which all
the T=0 matrix elements are set to a constant.  For the latter two
interactions the T=1 matrix elements are the same as for FPD6.  We are
thus able to study the effects of the fluctuating T=0 matrix elements
on M1 and GT rates.

\end{abstract}
\vspace{2.5in}

\newpage

\section{Introduction}

We have previously studied the effects of T=0 two body matrix elements
on energy levels. ~\cite{dow1,dow2} In this work we will focus on the
transition rates.  In the former work we used an FPD6 interaction to
get the energy levels of $^{44}$Ti, $^{46}$Ti, and $^{48}$Ti as well
as a second interaction wherein the T=0 matrix elements were set to
zero while the T=1 matrix elements we left unchanged.  In a single j
shell calculation of $^{44}$Ti we found that the energy levels of the
yrast even spins J=2-12 were very little affected by this apparently
severe change.  The odd spin T=0 states (not yet found experimentally)
were lowered in energy somewhat when this approximation was made.  In
this single j shell calculation many degeneracies appeared e.g. the
J=$9^{+}_1$ and 10$^{+}_1$ states.  The reason for these degeneracies
was explained in the first two references. ~\cite{dow1,dow2}

In a full fp calculation the even spin spectrum spread out a bit more -
leaning slightly toward a rotational spectrum when the T=0 matrix
elements were reintroduced (ie full FPD6 interaction) but only
slightly. It thus appeared that keeping only the T=1 matrix elements
led to a reasonable spectrum and the T=0 matrix elements were only
needed for fine tuning.

In the previous work we focused on excitation energies.  We now
examine the M1 and G-T transition strengths in the same nuclei to see
whether these strengths are more sensitive to T=0 matrix elements than
are the energy levels.  For completeness we also look at some M3
transition strengths.

\section{The Calculation}

Whereas in our previous works we considered only one modification of
the basis FPD6 interaction here we consider two.  We denote our three
interactions as follows:\\

Interaction A Set all T=0 two body matrix elements of FPD6 to zero;
keep all T=1 matrix elements of FPD6 unchanged. \\

Interaction B Set all T=0 two body matrix elements of FPD6 to a constant;
keep all T=1 matrix elements of FPD6 unchanged.\\

Interaction C Unmodified FPD6 interaction.\\

It should be mentioned that there is no difference in the results for
the \underline{spectrum} of the states of a given isospin in a single
j shell calculation between interaction A and B.  Of course the
ground state energy (binding energy) will be affected as will the
relative energies of states with different isospins.

However when configuration mixing is included there will be a
difference in the spectrum of states of a given isospin.  There is a
difference between setting the T=0 matrix elements equal to a constant
and introducing a constant T=0 interaction $c (\frac{1}{4} - t(1)
\cdot t(2))$.  With the latter there will be no change in the
\underline{spectrum} of states of a given isospin when we change the
value of c even in a large space calculation.  We get the same answer
whether c is positive, negative or zero (again the binding energy
\underline{will} be affected).

With the above constant T=0 interaction matrix elements of the form
$<[j_1,j_2]^{J,T=0}V[j_3,j_4]^{J,T=0}>$ will vanish if $(j_3,j_4) \ne
(j_1,j_2)$.  However for interaction B it will be a constant, the same constant as for the diagonal matrix elements.

We can regard the results for going from A to B to C as respectively studying the effects of

a) no T=0 interaction

b) An average T=0 interaction

c) fluctuations in the T=0 interaction with possible T=0 pairing

\section{Results}

In tables I-VI we give the summed strength B(M1) for the three
interactions A, B and C.  We display results for total B(M1),
B(M1)$_{spin}$, and B(M1)$_{orbital}$.  

The respective g factors are\\
B(M1);              g$_{s_{\pi}}$= 5.586,   g$_{s_{\nu}}$= -3.826,   g$_{l_{\pi}}$= 1,   g$_{l_{\nu}}$= 0 \\
B(M1)$_{spin}$;     g$_{s_{\pi}}$= 5.586,   g$_{s_{\nu}}$= -3.826,   g$_{l_{\pi}}$= 0,   g$_{l_{\nu}}$= 0 \\
B(M1)$_{orbital}$;  g$_{s_{\pi}}$= 0,       g$_{s_{\nu}}$= 0,        g$_{l_{\pi}}$= 1,   g$_{l_{\nu}}$= 0 \\

The six tables are as follows\\
I. $^{44}$Ti  J=0 T=0 $\rightarrow$ J=1 T=0\\
II. $^{44}$Ti  J=0 T=0 $\rightarrow$ J=1 T=1\\
III. $^{46}$Ti  J=0 T=1 $\rightarrow$ J=1 T=1\\
IV. $^{46}$Ti  J=0 T=1 $\rightarrow$ J=1 T=2\\
V. $^{48}$Ti  J=0 T=2 $\rightarrow$ J=1 T=2\\
VI. $^{48}$Ti  J=0 T=2 $\rightarrow$ J=1 T=3\\

The first case $^{44}$Ti J=0 T=0 $\rightarrow$ J=1 T=0 is atypical
because of the single j result that the M1 rates are zero.  This is
easily understood as arising from the fact that an isoscalar M1
operator $\vec{\mu}$ can be replaced by g$\vec{J}$ in a single j
shell, and the total angular momentum operator $\vec{J}$ cannot induce
M1 transitions.

In this table we introduce the parameter t which is the number of
nucleons excited from the f$_{7/2}$ shell (t should not be confused
with isospin T).  Thus t=0 corresponds to a single j shell calculation
(f$_{7/2}^{4}$ in $^{44}$Ti) while t=4 would correspond to all four
nucleons free to roam the entire f-p shell.  Since the B(M1)'s are
very small and this case atypical we shall not pursue a discussion of
it.

We next consider the transition J=0 T=0 $\rightarrow$ J=1 T=1 in
$^{44}$Ti.  We will now find a pattern of behavior more typical of
what happens in the other nuclei.  For the single j shell case (t=0)
interactions A and B give identical results.  This was explained
earlier.  Here we have performed calculations tuning t up to t=4 and
we discuss this calculation.

Comparing interactions A and C (for t=4) we find that the
reintroduction of the T=0 matrix elements causes the spin B(M1) to
decrease from 9.296 $\mu_n^2$ to 3.267 $\mu_n^2$.  The orbital B(M1)
increases by about a factor of two from 1.121 $\mu_n^2$ to 2.144
$\mu_n^2$.  These results are consistent with previous works where it
was noted that in the SU(4) limit the orbital B(M1) is large and the
spin B(M1) is zero.  The SU(4) limit is a case of high collectivity with
the other extreme being the single j shell limit.  It is clear that
reintroducing the T=0 matrix elements into the calculation will cause
nuclear collectivity to increase.

For the heavier nuclei we only go up to t=2 so it is instructive to
compare the t=2 and t=4 calculations in $^{44}$Ti.  We will focus on
interaction C.  Relative to t=1 we get a reduction in the t=2
calculation of B(M1)$_{spin}$ from 8.438 $\mu_n^2$ to 4.680 $\mu_n^2$.
When we go to t=4 the trend continues with B(M1)$_{spin}$ further
reduced to 3.267 $\mu_n^2$.  On the other hand the orbital strength
increases from 1.317 $\mu_n^2$ to 1.926 $\mu_n^2$ to 2.144 $\mu_n^2$
as we go from t=1 to t=2 to t=4.  This means excitation energies also
go steadily up.  These results are consistent with the fact that as we
increase the configuration mixing we increase the collectivity.  We
should keep in mind that for the heaviest titanium isotopes where we
limit calculations to t=2 we might be underestimating the collectivity
and the difference in t=2 and t=1 calculations would continue to grow
if we enlarged the space further.

\section{The nucleus $^{46}$Ti}

Since $^{44}$Ti is unstable no M1 excitation measurements have been
performed on this nucleus.  However the next nucleus that we consider
$^{46}$Ti has been extensively studied via inelastic scattering by the
Darmstadt group. ~\cite{dstadt}

We first consider the summed strength for the T=1 $\rightarrow$ T=1
M1 transitions.  We immediately see big changes as we go from
interaction A to C.  At the t=0 level (f$_{7/2 \pi}^{2}$f$_{7/2
\nu}^{4}$) the A and B interactions yield very small values for all
three M1's.  When the full interaction is reinstituted, the three
B(M1)'s all increase by a nearly constant factor of about 4.6.

In the largest space  calculation that we have done (t=2) B(M1) and
B(M1)$_{spin}$ increase from A to C but the most dramatic increase is
in the B(M1)$_{orbital}$.  The values there increasing from 0.1850 to
0.7095 $\mu_{N}^{2}$. This is almost a factor of four increase in the
orbital (scissors mode) strength.  So the T=0 matrix elements are
\underline{vital} for the enhanced B(M1).

One sees that most of this increase in orbital strength also occurs
with the B interaction.  This would suggest that it is mostly an
average T=0 effect rather than being due to a fluctuation in the matrix
elements or T=0 pairing.

Note that the B(M1)$_{spin}$ also gets some enhancement (4.996
$\rightarrow$ 5.537) but it is not so dramatic.  For the channel 0$_1$
$\rightarrow$ 1$_2$ the B(M1)$_{spin}$ get substantially quenched (3.941
$\rightarrow$ 1.717) as one goes from A to C, but again the orbital
summed strength gets enhanced (0.3846 $\rightarrow$  0.5138
$\mu _{N} ^{2}$).  Note also that the mean energies of the modes go up
substantially.

\section{The Nucleus $^{48}$Ti}

The behavior for $^{48}$Ti is similar to that of $^{46}$Ti.  In
going from the A to C interactions the values of B(M1)$_{orbital}$
increases substantially from 0.1931 to 0.5816 $\mu_{N}^{2}$ for J=0 T=2
$\rightarrow$ J=1 T=2 and from 0.1555 to 0.2719 $\mu_{N}^{2}$ for J=0
T=2 $\rightarrow$ J=1 T=3.

We still see the orbital enhancement is well described also by the
interaction B.  This implies again that it is mostly an average T=0
effect rather than being due to a fluctuation in the matrix elements
or T=0 pairing.

Note that in the single j shell calculation (t=0) the A and B
interactions give identical results for the three B(M1)'s but the mean
energies are different.  This is also true in $^{46}$Ti.  Only when
there is configuration mixing does the constant T=0 interactions A and
B differ as far as the B(M1)'s are concerned.

\section{Gamow-Teller Transition}

For the isovector B(M1) we have the relation 
\begin{equation}
\frac{B(GT)_{(T,-T)\rightarrow (T^{'}, -T+1)}}{B(M1)_{(T,T)\rightarrow
(T^{\prime},T)}} = 
constant \frac{\left( \begin{array}{ccc}
T^{\prime} & T & 1\\
T-1&-T& 1\\
\end{array} \right)^{2} } 
{ \left( \begin{array}{ccc}
T^{\prime} & T & 1\\
T&-T& 0\\
\end{array} \right)^{2} }
\end{equation}
$= 
 \left\{ \begin{array}{c}
constant (1) $ if T$^{\prime}=T$$ \\
  constant \frac{1}{(2T+1)} $  if 
T$^{\prime}=T+1$    $   \\
\end{array}\right\}
$.

For G-T there is one channel that is never present for M1's, J=0 T
$\rightarrow$ J=1$^{+}$ T-1.

We present results for GT transitions in tables VII to IX.  For
$^{46}$Ti we see even at the t=0 level a big change in the rate when
the full FPD6 is used for this channel.  (0.828 $\rightarrow$ 0.361).
At the t=2 level the change is from 4.666 to 2.033 more than a factor
of two reduction.

This large of a difference does not occur with the interaction B.
This suggests that for this channel pairing effects (alternatively
deviations from the average T=0 interaction) are important.

In tables X and XI we show the trends as the configuration space is
increased from t=1 to t=2.  We present the ratio of correspondence M1
and GT rates for [t=2]/[t=1].

We can see except for the anomolus T=0 $\rightarrow$ T=0 transition in
$^{44}$Ti, the obrital B(M1)'s get enhaced as the configuaration space
increased.  In all cases on the other hand, the spin B(M1)'s decrease
as the configuration space is increased.

Looking at the corresponding numbers for GT ratios, we note that
[t=2]/[t=1] ratios are the same as for the corresponding spin B(M1)'s
except for the T$_{f}$=T$_{i}$ case.In this case there is an isoscalar
contribution to B(M1) as well as the dominat isovector contribution.

There is a relationship between
$A= B(GT)_{(T,-T)} \rightarrow B(GT)_{(T+1),(-T+1)}$
and
$B=B(GT)_{(T,-T)} \rightarrow B(GT)_{(T+1),(-T-1)}$
i.e.
$B(GT)_{Ti\rightarrow V(T+1)}$ and $B(GT)_{Ti\rightarrow Sc(T+1)}$ 

The ratio A/B is equal to \\
$\frac{\left( \begin{array}{ccc}
(T+1) & T & 1\\
-(T+1)&T& 1\\
\end{array} \right)^{2}}{ \left( \begin{array}{ccc}
(T+1) & T & 1\\
(T-1)&-T& 1\\
\end{array} \right)^{2}}$=(2T+1)(2T+2)
Thus \\
$\frac{ B(GT)_{^{46}Ti\rightarrow ^{46}Sc(T=2)}            }
{    B(GT)_{^{46}Ti\rightarrow ^{46}V(T=2)}            }=6$
and
$\frac{    B(GT)_{^{48}Ti\rightarrow ^{48}Sc(T=3)}         }
{      B(GT)_{^{48}Ti\rightarrow ^{48}V(T=3)}        }=15$

In principle then one should get the 3(N-Z) sum rule without doing the
(n,p) reaction on $^{46}$Ti.  For example the $^{46}$Ti sum rule reads

$
B(GT)_{^{46}Ti\rightarrow ^{46}V(T=0)}
+B(GT)_{^{46}Ti\rightarrow ^{46}V(T=1)}
+B(GT)_{^{46}Ti\rightarrow ^{46}V(T=2)}
-B(GT)_{^{46}Ti\rightarrow ^{46}Sc(T=2)}
=3(N-Z)
$

We can write this as 
$
B(GT)_{^{46}Ti\rightarrow ^{46}V(T=0)}
+B(GT)_{^{46}Ti\rightarrow ^{46}V(T=1)}
-5 B(GT)_{^{46}Ti\rightarrow ^{46}V(T=2)}=3(N-Z)
$

However in practice it is very difficult if not impossible to seperate
the isospin components.

\section {Comments on M3 transitions}

In tables XII to XVII we present the results for M3 transitions from
the ground states of $^{44,46,48}$Ti.  The tables are presented in the
same format as those for M1's in Tables I thru VI.  One further note
is that in table XVI on the lowest 1000 states are considered in the
t=2 case for $^{48}$Ti.

The relative contribution of orbit to spin for B(M3) is much less than
for B(M1).

In general going form interaction A to C causes the orbital M3 to be
enhanced.  For the spin case these are mixed results sometimes there
is a quenching others an enhancement.

\begin{table}
\caption{Summed B(M1) strengths and mean excitation energies for $^{44}$Ti transition from J=0$^{+}$ T=0 to J=1$^{+}$ T=0.}
\begin{tabular}{ccccccc}
\tableline
 t=0 & A  & $\bar{E_{A}}$ &  B   &  $\bar{E_{B}}$   &  C  &  $\bar{E_{C}}$   \\
TOTAL&  &          &   &    &   &   \\
SPIN &  &          &   &    &   &   \\  
ORBIT&  &          &   &    &   &   \\
 t=1 & &  &   &     &   &    \\
TOTAL&0.0585 & 10.121    &0.05287 & 10.910   & 0.0547   & 9.246   \\
SPIN &0.3138  & 10.124    &0.2835  & 10.914   & 0.2936  &  9.298   \\  
ORBIT&0.1013  & 10.118    &0.09153 & 10.911   & 0.0948  &  9.279   \\
 t=2 &   &  &    &     &   &    \\
TOTAL&0.0325  & 11.563   & 0.03344 & 13.194   & 0.0274  & 11.422   \\
SPIN &0.1743  & 11.560   & 0.1793  & 13.196   & 0.1470  & 11.422   \\  
ORBIT&0.0563  & 11.564   & 0.0579  & 13.193   & 0.0475  & 11.422   \\
 t=4 &   &  &    &    &    &    \\
TOTAL& 0.0333  & 10.750   & 0.03130 & 12.990   & 0.0200   & 11.705   \\
SPIN & 0.1788  & 10.749   & 0.1679  & 12.990   & 0.1073   & 11.696  \\  
ORBIT& 0.0577 & 10.748   & 0.0542  & 12.989   & 0.0340   & 11.701  \\
\tableline
\end{tabular}
\end{table}
\begin{table}
\caption{Summed B(M1) strengths and mean excitation energies for $^{44}$Ti transition from J=0$^{+}$ T=0 to J=1$^{+}$ T=1.}
\begin{tabular}{ccccccc}
\tableline
 t=0 & A  & $\bar{E_{A}}$ &  B   &  $\bar{E_{B}}$   &  C  &  $\bar{E_{C}}$   \\
TOTAL& 5.955  & 3.249 & 5.955   & 4.248   & 4.442   & 6.002   \\
SPIN & 2.221  & 3.249 & 2.221   & 4.249   & 1.657   & 6.000   \\  
ORBIT& 0.9025 & 3.249 & 0.9025  & 4.429   & 0.6735  & 6.002   \\
 t=1 &   &  &     &     &    &    \\
TOTAL&12.540  & 6.200  &12.14 & 6.695   & 9.774   & 7.127    \\
SPIN &10.860  & 7.776  & 9.24 & 9.123   & 8.438   & 8.933    \\  
ORBIT&0.9919  & 4.063  & 1.601 & 5.712   & 1.313   & 8.149    \\
 t=2 &   &  &     &     &    &    \\
TOTAL&11.320  & 7.110 & 11.570  & 7.620  & 6.630  & 8.639  \\
SPIN & 9.100  & 8.964 & 8.252   &10.547  & 4.680  &10.589  \\  
ORBIT& 1.111  & 5.425 & 1.907   & 6.890  & 1.926  &10.249  \\
 t=4 &   &  &     &     &    &     \\
TOTAL&11.42  & 7.054  & 10.500  & 8.430   & 5.349   & 9.121   \\
SPIN & 9.296  & 8.772  & 7.336  &11.146   & 3.267   & 10.801  \\  
ORBIT& 1.121  & 5.615  & 2.084  & 7.654   & 2.144   & 10.881  \\
\tableline
\end{tabular}
\end{table}
\begin{table}
\caption{Summed B(M1) strengths and mean excitation energies for $^{46}$Ti transition from J=0$^{+}$ T=1 to J=1$^{+}$ T=1.}
\begin{tabular}{ccccccc}
\tableline
 t=0 & A  & $\bar{E_{A}}$ &  B   &  $\bar{E_{B}}$   &  C  &  $\bar{E_{C}}$   \\
TOTAL&0.2384  & 2.704 & 0.2384   & 2.704  & 1.092   & 3.380   \\
SPIN &0.08892  &2.704 & 0.0889   & 2.704  & 0.4073  & 3.381  \\  
ORBIT&0.03614  &2.704 & 0.03614  & 2.704  & 0.1655  & 3.381  \\
 t=1 &   &  &     &    &    &   \\
TOTAL& 7.004 & 8.905   &7.526 & 9.302 & 8.695   & 8.476  \\
SPIN & 7.707 & 9.323   &7.280 &10.204 & 7.793   & 9.523  \\  
ORBIT& 0.2504& 7.468   &0.4501& 7.447 & 0.5865  & 6.723  \\
 t=2 &   &  &     &     &    &     \\
TOTAL& 4.604  & 10.469 & 8.026   & 9.348    & 6.665   & 9.404   \\
SPIN & 4.996  & 10.880 & 6.840   &10.636    & 5.537   &10.466   \\  
ORBIT& 0.1850 & 8.914  &0.6701   & 7.593    &0.7095   & 9.257  \\
\tableline
\end{tabular}
\end{table}
\begin{table}
\caption{Summed B(M1) strengths and mean excitation energies for $^{46}$Ti transition from J=0$^{+}$ T=1 to J=1$^{+}$ T=2.}
\begin{tabular}{ccccccc}
\tableline
 t=0 & A  & $\bar{E_{A}}$ &  B   &  $\bar{E_{B}}$   &  C  &  $\bar{E_{C}}$   \\
TOTAL& 1.874   & 3.882  & 1.874  & 5.886   & 0.8320   & 9.006  \\
SPIN & 0.699  &  3.882  & 0.699  & 5.886   & 0.3103   & 9.007  \\  
ORBIT& 0.2841  & 3.882  & 0.284  & 5.886   & 0.1261   & 9.009  \\
 t=1 &   &  &     &     &    &     \\
TOTAL& 4.400  & 7.615 & 3.209 & 9.716 & 1.918 & 11.413 \\
SPIN & 4.487  & 8.703 & 3.297 & 11.732& 2.688 & 12.209 \\  
ORBIT& 0.2979 & 5.498 & 0.446 & 8.854 & 0.3153& 2.029  \\
 t=2 &   &  &    &    &   &     \\
TOTAL& 4.210 & 8.658 & 2.992 & 11.651 & 1.275 & 13.027   \\
SPIN & 3.941  &10.056& 3.196& 13.661 & 1.717  & 13.640   \\  
ORBIT& 0.3846 & 7.057& 0.508& 10.688 & 0.5138 & 14.296   \\
\tableline
\end{tabular}
\end{table}
\begin{table}
\caption{Summed B(M1) strengths and mean excitation energies for $^{48}$Ti transition from J=0$^{+}$ T=2 to J=1$^{+}$ T=2.}
\begin{tabular}{ccccccc}
\tableline
 t=0 & A  & $\bar{E_{A}}$ &  B   &  $\bar{E_{B}}$   &  C  &  $\bar{E_{C}}$   \\
TOTAL& 0.1679  & 2.905 & 0.1679   & 2.905    & 0.6021   & 3.640  \\
SPIN & 0.0623  &2.904 & 0.0623   & 2.904   & 0.2246  & 3.641   \\  
ORBIT& 0.0255 & 2.905 & 0.02545  & 2.905   & 0.0912 & 3.641   \\
 t=1 &   &  &     &     &    &    \\
TOTAL& 10.86  & 9.438 & 11.20  & 10.0625 & 11.81   & 9.941  \\
SPIN & 11.88  & 9.604   &11.09  & 10.586  & 11.54  &10.426     \\  
ORBIT& 0.2422  & 8.303  & 0.4486  &8.598   & 0.4484  & 8.669    \\
 t=2 &   &  &     &     &    &    \\
TOTAL& 7.666  & 10.968  &11.71   &10.487    &8.999   &10.871   \\
SPIN & 8.246  & 11.108  & 10.77  &11.151    & 8.393  & 11.315   \\  
ORBIT& 0.1931 & 10.005  & 0.6303  & 9.229   & 0.5816  &11.724   \\
\tableline
\end{tabular}
\end{table}
\begin{table}
\caption{Summed B(M1) strengths and mean excitation energies for $^{48}$Ti transition from J=0$^{+}$ T=2 to J=1$^{+}$ T=3.}
\begin{tabular}{ccccccc}
\tableline
 t=0 & A  & $\bar{E_{A}}$ &  B   &  $\bar{E_{B}}$   &  C  &  $\bar{E_{C}}$   \\
TOTAL&0.5454  & 4.499 & 0.5454   & 7.499    &0.1894   &11.705   \\
SPIN &0.2034  & 4.499 & 0.2304   & 7.498    &0.0707   &11.701   \\  
ORBIT&0.0827  & 4.499 & 0.08266  & 7.499    &0.0287   &11.700   \\
 t=1 &   &  &     &     &    &    \\
TOTAL&2.245  & 8.842    &1.276  &12.594   &0.8479  &13.587    \\
SPIN &2.695  & 9.301   &1.816  &13.607   &1.567  & 13.835   \\  
ORBIT&0.1089  &7.243    &0.1816  &12.153   &0.1793  &13.882    \\
 t=2 &   &  &     &     &    &    \\
TOTAL&2.108  &10.137  &1.345   &14.885    &0.5532   &15.454   \\
SPIN &2.356  &10.772  &1.945   &15.599    &1.042   & 15.134  \\  
ORBIT&0.1555  &9.256  &0.2374   &13.955    &0.2719   &16.775   \\
\tableline
\end{tabular}
\end{table}

\begin{table}
\caption{Gamow-Teller Transitions from the J=0$^{+}$ T=0 ground state of $^{44}$Ti to J=1$^{+}$ states.}
\begin{tabular}{ccccc}
\tableline
      & Final State   & A     & B     & C \\
$[t=0]$ & $^{44}$Sc T=1 & 1.315 & 1.315 & 0.9807\\
      & $^{44}$V  T=1 & 1.315 & 1.315 & 0.9807\\
$[t=1]$ & $^{44}$Sc T=1 & 6.430 & 5.470 &  4.995\\
      & $^{44}$V  T=1 & 6.430 & 5.470 &  4.995\\
$[t=2]$ & $^{44}$Sc T=1 & 5.387 & 4.886 &  2.771\\
      & $^{44}$V  T=1 & 5.387 & 4.886 &  2.771\\
$[t=4]$ & $^{44}$Sc T=1 & 5.503 & 5.470 &  1.934\\
      & $^{44}$V  T=1 & 5.503 & 5.470 &  1.934\\
\tableline
\end{tabular}
\end{table}

\begin{table}
\caption{Gamow-Teller Transitions from the J=0$^{+}$ T=1 ground state of $^{46}$Ti to J=1$^{+}$ states.}
\begin{tabular}{ccccc}
\tableline
      & Final State   & A     & B     & C \\
$[t=0]$ & $^{46}$Sc T=2 & 0.828 & 0.828 & 0.367 \\
      & $^{46}$V  T=2 & 0.139 & 0.138 & 0.061 \\
      & $^{46}$V  T=1 & 0.0526& 0.0526& 0.241 \\
      & $^{46}$V  T=0 & 4.661 & 4.661 &  4.089\\
$[t=1]$ & $^{46}$Sc T=2 &5.312& 3.904 & 3.185  \\
      & $^{46}$V  T=2 &0.8854  &0.6507 &0.5308  \\
      & $^{46}$V  T=1 &5.261 &4.892 &5.228  \\
      & $^{46}$V  T=0 &8.266 &6.694 &5.749  \\
$[t=2]$ & $^{46}$Sc T=2 &4.666  &3.784  &2.033   \\
      & $^{46}$V  T=2 &0.777  &0.6307  &0.3388  \\
      & $^{46}$V  T=1 &3.442  &4.505  & 3.734 \\
      & $^{46}$V  T=0 &9.254  &7.050  & 6.369 \\
\tableline
\end{tabular}
\end{table}

\begin{table}
\caption{Gamow-Teller Transitions from the J=0$^{+}$ T=2 ground state of $^{48}$Ti to J=1$^{+}$ states.}
\begin{tabular}{ccccc}
\tableline
      & Final State   & A     & B     & C \\
$[t=0]$ & $^{48}$Sc T=3 &0.3612 &0.3612 &0.1255 \\
      & $^{48}$V  T=3 &0.02408  &0.02408&0.008365 \\
      & $^{48}$V  T=2 &0.01854  &0.01854&0.06647  \\
      & $^{48}$V  T=1 &8.367    &8.367  &8.099 \\
$[t=1]$ & $^{48}$Sc T=3 &4.787  &3.226  &2.783  \\
      & $^{48}$V  T=3 &0.3191   &0.2150 &1.855 \\
      & $^{48}$V  T=2 &4.288    &3.942  &4.094 \\
      & $^{48}$V  T=1 &18.29    &15.82  &15.40 \\
$[t=2]$ & $^{48}$Sc T=3 &4.184 &3.454 &1.851 \\
      & $^{48}$V  T=3 &0.279 &0.230 &0.1234 \\
      & $^{48}$V  T=2 &3.019 &3.738 &3.041 \\
      & $^{48}$V  T=1 &18.37 &16.48 &15.66 \\
\tableline
\end{tabular}
\end{table}

\begin{table}
\caption{Ratios of M1's in different model spaces in FPD6}
\begin{tabular}{cccc}
\tableline
          & Total & Spin & Orbital\\
$^{44}$Ti &       &      &        \\
0$_{0}$ $\rightarrow$ 1$_{0}$ (t=2/t=1) & 0.501 & 0.501 & 0.501 \\
0$_{0}$ $\rightarrow$ 1$_{0}$ (t=4/t=2) & 0.366 & 0.365 & 0.359 \\
0$_{0}$ $\rightarrow$ 1$_{0}$ (t=2/t=1) & 0.678 & 0.554 & 1.467 \\
0$_{0}$ $\rightarrow$ 1$_{0}$ (t=4/t=2) & 0.551 & 0.387 & 1.633 \\
$^{46}$Ti &  (t=2/t=1)      &      &        \\
0$_{1}$ $\rightarrow$ 1$_{1}$ & 0.770 & 0.771 & 1.210 \\
0$_{1}$ $\rightarrow$ 1$_{2}$ & 0.665 & 0.639 & 1.630 \\
$^{48}$Ti &  (t=2/t=1)      &      &        \\
0$_{2}$ $\rightarrow$ 1$_{2}$ & 0.761 & 0.727 & 1.297 \\
0$_{2}$ $\rightarrow$ 1$_{3}$ & 1.006 & 0.665 & 1.587 \\
\tableline
\end{tabular}
\end{table}

\begin{table}
\caption{Ratio ($t=2 / t=1$) of GT's in different model spaces in FPD6}
\begin{tabular}{ccc}
\tableline
$^{44}$Ti &       &      \\
          & $^{44}$Sc T=1 & 0.555 \\
          & $^{44}$V T=1  & 0.555 \\
$^{46}$Ti &       &        \\
          &$^{46}$Ti T=2 & 0.638 \\
          &$^{46}$V T=2 & 0.638 \\
          &$^{46}$V T=1 & 0.714 \\
          &$^{46}$V T=0 & 1.107 \\
$^{48}$Ti &       &        \\
          &$^{48}$Ti T=3 & 0.665 \\
          &$^{48}$V T=3 & 0.665 \\
          &$^{48}$V T=2 & 0.743 \\
          &$^{48}$V T=1 & 1.017 \\
\tableline
\end{tabular}
\end{table}

\begin{table}
\caption{Summed B(M3) strengths and mean excitation energies for $^{44}$Ti transition from J=0$^{+}$ T=0 to J=3$^{+}$ T=0.}
\begin{tabular}{ccccccc}
\tableline
 t=0 & A  & $\bar{E_{A}}$ &  B   &  $\bar{E_{B}}$   &  C  &  $\bar{E_{C}}$   \\
TOTAL& 53.07  & 3.852  & 53.95   & 3.852    & 223.8   & 4.821   \\
SPIN & 11.63  & 3.852  & 11.82   & 3.853    & 49.03   & 4.822   \\  
ORBIT& 15.02  & 3.852  & 15.26   & 3.854    & 63.31   & 4.821   \\
 t=1 &   &  &     &     &    &    \\
TOTAL&1124  &5.588    &1295  & 6.855   & 1146  &6.889    \\
SPIN &1102  &8.009    &1051  & 8.924   & 1046  &9.152   \\  
ORBIT&375.5  &7.671    &410.6& 8.327   &  369.8&8.077    \\
 t=2 &   &  &     &     &    &     \\
TOTAL&529.1  &7.171  &1057   & 8.248   & 724   & 9.362   \\
SPIN &578  &9.666  &686.4   &10.943  & 603.1  &11.438   \\  
ORBIT&201.1&9.364  &341.3   &10.056  & 230.7  &10.152   \\
 t=4 &   &  &     &     &    &    \\
TOTAL& 510.4  &6.549  &1179    & 8.736    & 669.9  & 9.903 \\
SPIN & 573.0  &8.979  &663.7   &10.866    & 487.9  &11.898 \\  
ORBIT& 204.5  &8.792  &403.1   & 10.794   & 242.8  &10.613 \\
\tableline
\end{tabular}
\end{table}
\begin{table}
\caption{Summed B(M3) strengths and mean excitation energies for $^{44}$Ti transition from J=0$^{+}$ T=0 to J=3$^{+}$ T=1.}
\begin{tabular}{ccccccc}
\tableline
 t=0 & A  & $\bar{E_{A}}$ &  B   &  $\bar{E_{B}}$   &  C  &  $\bar{E_{C}}$   \\
TOTAL&12890  &2.991& 12890  & 3.993   &15120   & 5.206   \\
SPIN &8770  &2.991 & 8770   & 3.993   &10290   & 5.204   \\  
ORBIT&396  &2.990  & 396    & 3.990   & 464.4  & 5.207   \\
 t=1 &   &  &     &     &    &    \\
TOTAL& 41730  & 5.722   &32910  & 6.879   & 37980  & 7.254   \\
SPIN & 39380  & 6.402   &31380  & 7.597   & 35120  & 7.921   \\  
ORBIT& 761.6  & 5.469   &738.8  & 7.755   & 891.2  & 8.710   \\
 t=2 &   &  &     &     &    &     \\
TOTAL&33830  &6.908  &22640   & 9.090   &25640  & 8.982   \\
SPIN &31670  &7.629  &22240   & 9.834   &23400  & 9.705   \\  
ORBIT&670.4  &6.595  &701.7   & 9.662   &924.1  & 10.908  \\
 t=4 &   &  &     &     &    &     \\
TOTAL&33920  &6.701  &16500   &10.745    &19150   & 9.629   \\
SPIN &31960  &7.356  &16700   &11.371    &17600   & 10.250  \\  
ORBIT&679.1  &6.579  &760.6   &9.979     &982.5   & 11.63  \\
\tableline
\end{tabular}
\end{table}

\begin{table}
\caption{Summed B(M3) strengths and mean excitation energies for $^{46}$Ti transition from J=0$^{+}$ T=1 to J=3$^{+}$ T=1.}
\begin{tabular}{ccccccc}
\tableline
 t=0 & A  & $\bar{E_{A}}$ &  B   &  $\bar{E_{B}}$   &  C  &  $\bar{E_{C}}$   \\
TOTAL& 679.7  & 3.260  & 679.7   & 3.260   & 2896   & 4.043   \\
SPIN & 434.8  & 3.229  & 434.8   & 3.229   & 1849   & 4.019   \\  
ORBIT& 35.90  & 3.482  & 35.90   & 3.482   & 155.2  & 4.214   \\
 t=1 &   &  &    &     &    &     \\
TOTAL& 27220 & 7.351   & 26350  & 8.387    & 26830  & 8.233   \\
SPIN & 27520 & 7.751   & 25750  & 8.777    & 25930  & 8.589   \\  
ORBIT& 610   & 7.836   & 675.8  & 8.866    & 724.8  & 8.469   \\
 t=2 &   &  &    &     &    &     \\
TOTAL& 17030 & 9.065  &21280   & 9.671    & 18980   & 9.768   \\
SPIN & 17390 & 9.413  &20560   &10.083    & 18180   & 10.071  \\  
ORBIT& 371.3 & 9.499  &602.8   &10.567    &  592.1    & 10.758  \\
\tableline
\end{tabular}
\end{table}
\begin{table}
\caption{Summed B(M3) strengths and mean excitation energies for $^{46}$Ti transition from J=0$^{+}$ T=1 to J=3$^{+}$ T=2.}
\begin{tabular}{ccccccc}
\tableline
 t=0 & A  & $\bar{E_{A}}$ &  B   &  $\bar{E_{B}}$   &  C  &  $\bar{E_{C}}$   \\
TOTAL& 5354  &3.340  &5354   & 5.340   & 5983   & 6.936   \\
SPIN & 3642  &3.339  &3642   & 5.340   & 4070   & 6.936  \\  
ORBIT&164.5  &3.337  &164.5  & 5.337   & 183.8  & 6.937  \\
 t=1 &   &  &     &     &    &    \\
TOTAL&18850  &6.488  &14290  & 8.579   & 16090 & 9.627   \\
SPIN &18080  &7.151  &13850  & 9.357   & 15010 &10.380   \\  
ORBIT&346.2  &6.043  &348.7  & 9.415   & 406.6 &10.592    \\
 t=2 &   &  &     &     &    &     \\
TOTAL&16240  &7.820  &10850   &10.876  & 12310  &11.282   \\
SPIN &15490  &8.528  &10980   &11.685  & 11550  &12.104   \\  
ORBIT&324.3  &7.431  &343.2   &11.468  & 420.2  &12.848   \\
\tableline
\end{tabular}
\end{table}
\begin{table}
\caption{Summed B(M3) strengths and mean excitation energies for $^{48}$Ti transition from J=0$^{+}$ T=2 to J=3$^{+}$ T=2.}
\begin{tabular}{ccccccc}
\tableline
 t=0 & A  & $\bar{E_{A}}$ &  B   &  $\bar{E_{B}}$   &  C  &  $\bar{E_{C}}$   \\
TOTAL& 580.6 &3.340  &580.6   &3.340    & 2233   & 4.378   \\
SPIN & 376.9 &3.317  &376.8   &3.317    & 1445   & 4.374  \\  
ORBIT& 27.70 &3.516  &27.70   &3.516    & 108.7  & 4.419  \\
 t=1 &   &  &     &     &    &     \\
TOTAL& 43350  &7.645    &42340  &8.618    &42890  & 8.631    \\
SPIN &43780  & 7.947   &41210  &8.974    &41720  & 8.902  \\  
ORBIT&686.3  & 8.104   &796.7  &9.041    &807.9  & 8.791   \\
 t=2 &   &  &     &     &    &     \\
TOTAL&30100  & 9.236  & 37130   & 9.747    & 32780  & 9.826  \\
SPIN &30530  & 9.489  & 36140   & 10.086    & 31750  &10.04   \\  
ORBIT&431.9  & 9.771  & 705.6   & 10.916    & 649.1  & 11.132   \\
\tableline
\end{tabular}
\end{table}
\begin{table}
\caption{Summed B(M3) strengths and mean excitation energies for $^{48}$Ti transition from J=0$^{+}$ T=2 to J=3$^{+}$ T=3.}
\begin{tabular}{ccccccc}
\tableline
 t=0 & A  & $\bar{E_{A}}$ &  B   &  $\bar{E_{B}}$   &  C  &  $\bar{E_{C}}$   \\
TOTAL&1932  &3.715  & 1932   & 6.718    & 1897   & 8.809   \\
SPIN &1314  &3.716  & 1314  & 6.718   & 1290  & 8.814  \\  
ORBIT&59.35  &3.715  &59.35   & 6.716    & 58.27  & 8.809   \\
 t=1 &   &  &     &     &    &    \\
TOTAL&10570  &7.687    & 7110  & 11.049   & 7836  & 12.564   \\
SPIN &10710  &8.196    & 7543  & 11.644   &8074  & 13.141  \\  
ORBIT&174.6  &7.268    &170.3  &11.973    &188  & 13.511   \\
 t=2 &   &  &     &     &    &    \\
TOTAL&9333  &9.103  &5780   & 13.734   &6250   & 14.211  \\
SPIN &9381  &9.640  &6530   & 14.158   & 6524   & 14.785  \\  
ORBIT&171  &8.860  &180.5   &14.177    & 205.3  & 15.831   \\
\tableline
\end{tabular}
\end{table}

\end{document}